\documentclass[twoside,12pt]{article} 

\usepackage{amssymb}

\def\bbbn{{\rm I\!N}}

\def\bbbz{{{\hbox{\sf Z\kern-0.4em Z}}}}
\def\var{{\it var}}
\def\reduc{{\it reduc}}
\def\expl{{\it expl}}
\def\ptrule#1#2#3{\begin{tabular}[t]{cl}
{#2} & \\
\hrulefill & \hspace*{-1em}{\footnotesize {#1}}\\
{#3} & \\
\end{tabular}}
\def\ptrules#1#2#3#4#5{\begin{tabular}[t]{cl}
{#3} & \\
\hrulefill & \hspace*{-1em}{\footnotesize {#1}}\\
{#4} & \\
\hrulefill & \hspace*{-1em}{\footnotesize {#2}}\\
{#5} & \\
\end{tabular}}

\newtheorem{defth}{Definition}
\newenvironment{definition}{\begin{defth}\parindent=0mm}{\end{defth}}
\newtheorem{lemmath}{Lemma}
\newenvironment{lemma}{\begin{lemmath}\parindent=0mm}{\end{lemmath}}
\newtheorem{corollaryth}{Corollary}

\newtheorem{theoremth}{Theorem}
\newenvironment{theorem}{\begin{theoremth}\parindent=0mm}{\end{theoremth}}
\newenvironment{proof}{\begin{quotation}\noindent\parindent=0mm{\it Proof. }%
\begin{rm}}{\hfill$\square$\end{rm}\end{quotation}}
\newtheorem{exampleth}{\em Example}
\newenvironment{example}[1]{\begin{exampleth}{#1}\parindent=0mm\par\begin{rm}}%
{\end{rm}\end{exampleth}}

{\end{array}\end{displaymath}}

\begin{document} 
\pagestyle{myheadings} 
\markboth{AADEBUG 2000}{Value Withdrawal Explanation in CSP} 
\title{Value Withdrawal Explanation in CSP
\footnote{In M. Ducass\'e (ed), proceedings of the Fourth
International Workshop on Automated Debugging (AADEBUG 2000), August
2000, Munich. COmputer Research Repository (http://www.acm.org/corr/),
cs.SE/0012005; whole proceedings: cs.SE/0010035.}} 
\author{G\'erard Ferrand \and Willy Lesaint \and Alexandre Tessier\\
        LIFO, BP 6759, 45067 Orl\'eans Cedex 2, France\\ 
        http://www.univ-orleans.fr/SCIENCES/LIFO} 
\date{} 
\maketitle 
\begin{abstract} 
This work is devoted to constraint solving motivated by the debugging
of constraint logic programs a la GNU-Prolog. The paper focuses only
on the constraints. In this framework, constraint solving amounts to
domain reduction. A computation is formalized by a chaotic
iteration. The computed result is described as a closure. This model
is well suited to the design of debugging notions and tools, for
example failure explanations or error diagnosis. In this paper we
detail an application of the model to an explanation of a value
withdrawal in a domain. Some other works have already shown the
interest of such a notion of explanation not only for failure
analysis.
\end{abstract}

\section{Introduction}
\label{Sect:Introduction}

Constraint Logic Programming (CLP) \cite{MarStu-book-98} can be viewed
as the reunion of two programming paradigms : logic programming and
constraint programming. Declarative debugging of constraints logic
programs has been treated in previous works \cite{Tessier-aadebug-97}
and tools have been produced for this aim \cite{Tessier-discipl-00}
during the DiSCiPl (Debugging Systems for Constraint Programming)
ESPRIT Project. But these works deal with the clausal aspects of
CLP. This paper focuses on the constraint level alone. The tools used at
this level strongly depend on the constraint domain and the way to
solve constraints. Here we are interested in a wide field of
applications of constraint programming: {\em finite domains} and
{\em propagation}.

The aim of constraint programming is to solve Constraint Satisfaction
Problems (CSP) \cite{Tsang-book-93}, that is to provide an
instantiation of the variables which is correct with respect to the
constraints.

The solver goes towards the solutions combining two different methods.
The first one (labeling) consists in partitioning the domains until to
obtain singletons and, testing them. The second one (domain reduction)
reduces the domains eliminating some values which cannot be correct
according to the constraints. Labeling provides exact solutions
whereas domain reduction simply approximates them.  In general, the
labeling alone is very expensive and a good combination of the two
methods is more efficient.
In this paper labeling is not really treated. We consider only one
branch of the search tree: the labeling part is seen as additional
constraint to the CSP. In future work, we plan to extend our framework
in order to fully take into account labeling and the whole search tree
(instead of a single branch).

This kind of computation is not easy to debug because CSP are not
algorithmic programs \cite{Meier-cp-1995}. The constraints are reinvoked
according to the domain reductions until a fix-point is reached. But
the order of invocation is not known a priori.

The main contribution of this paper is to formalize the domain
reduction in order to provide a notion of explanation for the basic
event which is ``the withdrawal of a value from a domain''.
This notion of explanation is essential for the debugging of CSP
programs. Indeed, the disappearance of a value from a domain may be a
symptom of an error in the program. But the error is not always where the
value has disappeared and an analysis of the explanation of the value
withdrawal is necessary to locate the error.
\cite{GouBen-ase-99} provides a tool to find symptoms, this paper
provides a tool which could be used to find errors from symptoms.
Explanations are a tool to help debugging: we extract from a (wide)
computation a structured part (the explanation) which will be analyzed
more efficiently.

We are inspired by a constraint programming language over finite domains,
GNU-Prolog \cite{CodDia-jlp-96}, because its glass-box approach allows
a good understanding of the links between the constraints and the
rules.

To be easily understandable, the notion of explanation will be first
defined in a framework which includes arc consistency and next in a
more general framework which includes hyper-arc consistency and also
some weaker consistencies usually used in the implemented constraint
solvers.

An explanation is a subset of rules used during the computation and
which are responsible for the removal of a value from a domain.
Several works shown that detailed analysis of explanations have a lot
of applications \cite{GueJusPri-ejor-00,Jussien-phd-97}. In dynamic
problems, the explanations allow to retract constraints without
beginning the computation again. In backtracking algorithms, the
explanations avoid to repeatedly perform the same search work. This
intelligent backtracking can be applied to scheduling problems. It has
been proved efficient for Open-shop applications.  They are useful for
over-constrained problems too. Explanations provide a set of
constraints which can be relaxed in order to obtain a solution. But
these applications of explanations are outside the scope of this paper
(see \cite{Jussien-phd-97}). Here, our definitions of explanations are
motivated by applications to debugging, in particular to error
diagnosis.

An aspect of the debugging of constraint programs is to understand why
we have a failure (i.e. we do not obtain any solution)
\cite{Fabris-aadebug-97}. This case appears when a domain becomes
empty, that is no value of the domain belongs to a solution. So, an
explanation of why these values have disappeared provides an
explanation of the failure.

Another aspect is error diagnosis. Let us assume an expected semantics
for the CSP.  Consider we are waiting for a solution containing a
certain value for a variable, but this value does not appear in the
final domain.  An explanation of the value withdrawal help us to find
what is wrong in our program.
It is important to note that the error is not always the constraint
responsible of the value withdrawal. Another constraint may have made
a wrong reduction of another domain which has finally produced the
withdrawal of the value.
The explanation is a structured object in which this information may
be founded.

The paper is organized as follows. Section~\ref{Sect:Preliminaries}
gives some notations and basic definitions for Constraint Satisfaction
Problems. Section~\ref{Sect:Model} describes a model for domain
reduction. Section~\ref{Sect:Explanations} applies the model to
explanations. Next section is a conclusion.

\section{Preliminaries}
\label{Sect:Preliminaries}

We use the following notations: If $F=(F_i)_{i \in I}$ is a family
indexed by $I$, and $J \subseteq I$, we denote by $F|_J$ the family
$(F_j)_{j \in J}$ indexed by $J$. If $F=(F_i)_{i \in I}$ is a family
of sets indexed by $I$, we denote by $\prod F$ the product $\prod_{i
\in I} F_i = \{ (e_i)_{i \in I} \mid \mbox{for each }i \in I, e_i \in
F_i \}$.

Notations in terms of families and tuples as in \cite{Tsang-book-93}
are convenient notations in our framework.
They are more readable than notations in terms of cartesian products
as in \cite{Apt-tcs-99} for example.

\medskip

Here we only consider the framework of domain reduction as in
\cite{VanHentenryck-book-89,CodDia-jlp-96,Benhamou-alp-96,VanEmden-misc-95}.
More general framework is described in \cite{MonRos-ai-91}.

A {\em Constraint Satisfaction Problem} (CSP) is made of two parts,
the syntactic part:
\begin{itemize}
\item a finite set of variable symbols (variables in short) $V$;
\item a finite set of constraint symbols (constraints in short) $C$;
\item a function $\var : C \rightarrow {\cal P}(V)$, which associates
with each constraint symbol the set of variables of the constraint;
\end{itemize}
and the semantic part:
\begin{itemize}
\item a family of non empty domains indexed by the set of variables
$D=(D_x)_{x \in V}$, each $D_x$ is the domain of the variable denoted
by $x$ ($D_x \not= \emptyset$);
\item a family of relations (sets of tuples) $T = (T_c)_{c \in C}$ indexed by the set of constraints $C$, where for each $c \in C$, $T_c \subseteq
\prod D|_{\var(c)}$, the members of $T_c$ are called the {\em solutions} of
$c$.
\end{itemize}

A tuple $t \in \prod D$ is a {\em solution} of the CSP
$(V,C,\var,D,T)$ if for each $c \in C$, $t|_{\var(c)} \in T_c$.

We introduce some useful notations:
${\cal D} = \prod_{x \in V} {\cal P}(D_x)$ (the search space) and
${\cal D}(W) = \prod_{x \in W} {\cal P}(D_x)$.

For a given CSP, one is interested in the computation of the
solutions. The simplest method consists in generating all the tuples
from the initial domains, then testing them. This {\em generate and
test} method is clearly expensive for wide domains. So, one prefers to
reduce the domains first (``test'' and generate).

Here, we focus on the reduction stage. The computing domains must
contain all the solutions and must be as small as possible. So, these
domains are ``approximations'' of the set of solutions. We describe
now, a model for the computation of such approximations.

\section{A Model of the Operational Semantics}
\label{Sect:Model}

We consider a fixed CSP $(V,C,\var,D,T)$.

We propose here a model of the operational semantics of the
computation of approximations which will be well suited to define
explanations of basic events useful for debugging. Moreover main
classical results \cite{Apt-tcs-99,MonRos-ai-91} are proved again in
this model.

The goal is to compute an approximation of the solutions.
A way to achieve this goal is to associate with the constraints some
{\em reduction rules}.
A rule works on a subset of the variables of the CSP. It eliminates
from one domain (and only one in our framework based on hyper-arc
consistency) some values which are inconsistent with respect to the
other domains.

\begin{definition}
A {\em reduction rule} $r$ of type $(W,y)$, where $W \subseteq V$ and
$y \in W$, is a function $r : {\cal D}(W) \rightarrow {\cal P}(D_y)$
such that: for each $d, d' \in {\cal D}(W)$,
\begin{itemize}
\item {\em (monotonicity)} $(\mbox{for each } x \in W$, $d_x \subseteq d'_x)
\Rightarrow r(d) \subseteq r(d')$;
\item {\em (contractance)} $r(d) \subseteq d_y$.
\end{itemize}
\end{definition}

The solver is described by a set of rules associated with the
constraints of the CSP. We can choose more or less accurate rules for
each constraint (in general, the more accurate are the rules, the more
expensive is the computation).

Other works consider more general kinds of rules
\cite{Benhamou-alp-96,Apt-tcs-99}, their types have the form $(W, Z)$
with $Z \subseteq W \subseteq V$.

\begin{example}{Hyper-arc consistency}
Let $W \subseteq V$, $y \in W$, $T \subseteq \prod D|_W$ and $d \in
{\cal D}(W)$. The reduction rule $r$ of type $(W,y)$
defined by $r(d)=\{t_y \mid t \in (\prod d) \cap T\}$ is an hyper-arc
consistency rule. $r$ removes inconsistent values with respect to the
variable domains.

When $W$ is $\{x,y\}$ it is the well known {\em arc consistency} framework.
\end{example}

\begin{example}{GNU-Prolog}
In GNU-Prolog, such rules are written $x~in~r$ \cite{CodDia-jlp-96},
where $r$ is a range dependent on domains of a set of variables. The
rule {\tt x~in~0..max(y)} of type $(\{x,y\},x)$ is the function which
computes the intersection between the current domain of $x$ and the domain
$\{0,1,\ldots,max(y)\}$ where $max(y)$ is the greatest value in the
domain of $y$.
\end{example}

For the sake of simplicity, for each rule, we define its associated
reduction operator. This operator applies to the whole family of
domains. A single domain is modified: the domain reduced by the
reduction rule.

The {\em reduction operator} associated with the rule $r$ of type
$(W,y)$ is $\reduc_r : {\cal D} \rightarrow {\cal D}$ defined by: for
each $d \in {\cal D}$,
\begin{itemize}
\item
$\reduc_r(d)|_{V \setminus \{y\}} = d|_{V \setminus \{y\}}$;
\item
$\reduc_r(d)_y = r(d|_W)$.
\end{itemize}
Note that reduction operators are monotonic and contractant (but they
are not necessarily idempotent).

A reduction rule $r$ is {\em correct} if, for each $d \in {\cal D}$,
for each solution $t \in \prod D$,
$t \in \prod d \Rightarrow t \in \prod \reduc_r(d)$.

\begin{lemma}
A reduction rule $r$ of type $(W,y)$ is correct if and only if, for
each solution $t$, $r((\{t_x\})_{x \in W}) = \{t_y\}$.
\end{lemma}

\begin{proof}
$\Rightarrow$: apply the definition with $d$ ``reduced'' to a
solution.

$\Leftarrow$: because reduction operators are monotonic.
\end{proof}

In practice, each constraint of the CSP is implemented by a set of
reduction rules.

Let $c \in C$. A reduction rule $r$ of type $(W,y)$ with $W \subseteq
\var(c)$ is {\em correct} with respect to $c$ if, for each $d \in
{\cal D}$, for each $t \in T_c$, $t \in \prod
d|_{\var(c)} \Rightarrow t \in \prod \reduc_r(d)|_{\var(c)}$.

\begin{lemma}
A reduction rule $r$ of type $(W,y)$ is correct w.r.t. a constraint
$c$ if and only if, for each $t \in T_c$, $r((\{t_x\})_{x \in W}) =
\{t_y\}$.
\end{lemma}

\begin{proof}
$\Rightarrow$: apply the definition with $d=(\{t_x\})_{x \in V}$ such
that $(t_x)_{x \in \var(c)} \in T_c$.

$\Leftarrow$: because reduction operators are monotonic.
\end{proof}

Note that if a reduction rule $r$ is correct w.r.t. a constraint $c$
of the CSP then $r$ is correct. But the converse does not hold.

\begin{example}{GNU-Prolog}
The rule $r$ : {\tt x~in~0..max(y)} is correct with respect to the
constraint $c$ defined by $\var(c)=\{x,y\}$ and $T_c=\{(x \mapsto 0,y
\mapsto 0),(x \mapsto 0,y \mapsto 1),(x \mapsto 1,y \mapsto 1)\}$
($D_x = D_y = \{0,1\}$ and $c$ is the constraint $x \leq y$).
Indeed,
\begin{itemize}
\item $r(x \mapsto \{0\},y \mapsto \{0\})=\{0\} \cap \{0\}=\{0\}$;
\item $r(x \mapsto \{0\},y \mapsto \{1\})=\{0\} \cap \{0,1\}=\{0\}$;
\item $r(x \mapsto \{1\},y \mapsto \{1\})=\{1\} \cap \{0,1\} =\{1\}$.
\end{itemize}
\end{example}

Let $R$ be a set of reduction rules.

Intuitively, the solver applies the rules one by one replacing the
domains of the variables with those it computes. The computation stops
when one domain becomes empty (in this case, there is no solution), or
when the rules cannot reduce domains anymore (a common fix-point is reached).

We will show that if no reduction rule is ``forgotten'', the resulting
domains are the same whatever the order the rules are used.

The computation starts from $D$ and tries to reduce as much as
possible the domain of each variable using the reduction rules.

The {\em downward closure} of $D$ by the set of reduction rules $R$ is
the greatest common fix-point of the reduction operators associated
with the reduction rules of $R$.

The downward closure is the most accurate family of domains which can
be computed using a set of correct rules.  Obviously, each solution
belongs to this family.

\medskip

Now, for each $x \in V$, the inclusion over ${\cal P}(D_x)$ is assumed
to be a well-founded ordering (i.e. each $D_x$ is finite).

There exists at least two ways to compute the downward closure of $D$
by a set of reduction rules $R$:
\begin{enumerate}
\item
the first one is to iterate the operator
${\cal D} \rightarrow {\cal D}$ defined by
$d \mapsto (\bigcap_{r \in R} \reduc_r(d)_x)_{x \in V}$ from $D$ until to
reach a fix-point;
\item
the second one is the {\em chaotic iteration} that we are going to
recall.
\end{enumerate}

The following definition is inspired from Apt \cite{Apt-tcs-99}.

A {\em run} is an infinite sequence of operators of $R$. A run is {\em
fair} if each $r \in R$ appears in it infinitely often. Let us define
an {\em iteration} of a set of rules w.r.t. a run.

\begin{definition}
The {\em iteration} of the set of reduction rules $R$ from the domain
$d \in {\cal D}$ with respect to the run $r_1,
r_2, \dots$ is the infinite sequence $d^0, d^1, d^2, \dots$ defined
inductively by:
\begin{enumerate}
\item $d^0 = d$;
\item for each $j \in \bbbn$, $d^{j+1} = reduc_{r_{j+1}}(d^j)$.
\end{enumerate}
A {\em chaotic iteration} is an iteration w.r.t. a fair run.
\end{definition}

The operator $d \mapsto (\bigcap_{r \in R} \reduc_r(d)_x)_{x \in V}$
may reduce several domains at each step. But the computations are more
intricate and some can be useless. In practice chaotic iterations are
preferred, they proceed by elementary steps, reducing only one domain
at each step. The next result of confluence
\cite{CouCou-saipl-77} ensure that any chaotic iteration reaches the
closure. Note that, because $D$ is a family of finite domains, every
iteration from $D$ is stationary.

\begin{lemma}\label{lemma:chaotic iteration}
The limit of every chaotic iteration of the reduction rules $R$ from
$D$ is the {\em downward closure} of $D$ by $R$.
\end{lemma}

\begin{proof}
Let $\Theta$ be the downward closure of $D$ by $R$. Let $d^0, d^1,
\dots$ be a chaotic iteration of $R$ from $D$ with respect to $r_1,
r_2, \dots$. Let $d^{\omega}$ be the limit of the chaotic iteration.
Let $(A_i)_{i \in I} \sqsubseteq (B_i)_{i \in I}$ denotes: for each $i
\in I$, $A_i \subseteq B_i$.

For each $i$, $\Theta \sqsubseteq d^i$, by induction: $\Theta
\sqsubseteq d^0 = D$. Assume $\Theta \sqsubseteq d^i$, by
monotonicity, $reduc_{r_{i+1}}(\Theta) = \Theta \sqsubseteq
\reduc_{r_{i+1}}(d^i) = d^{i+1}$.

$d^{\omega} \sqsubseteq \Theta$: There exists $k \in \bbbn$ such that
$d^{\omega}=d^k$ because $\sqsubseteq$ is a well-founded ordering. The
run is fair, hence $d^k$ is a common fix-point of the reduction
operators, thus $d^k \sqsubseteq \Theta$ (the greatest common
fix-point).
\end{proof}

The fairness of runs is a convenient theoretical notion to state the
previous lemma. Every chaotic iteration stabilizes, so in practice the
computation ends when a common fix-point is reached. Moreover,
implementations of solvers use various strategies in order to
determinate the order of invocation of the rules.

In practice, when a domain becomes empty, we know that there is no solution,
so an optimization consists in stopping the computation before the
closure is reached. In that case, we say that we have a {\em failure
iteration}.

\section{Application to Event Explanations}
\label{Sect:Explanations}

Sometimes, when a domain becomes empty or just when a value is removed
from a domain, the user wants an explanation of this phenomenon
\cite{Jussien-phd-97,Fabris-aadebug-97}.  The case of failure is the
particular case where all the values are removed. It is the reason why
the basic event here will be a value withdrawal. Let us consider an
iteration, and let us assume that at a step a value is removed from
the domain of a variable. In general, all the rules used from the
beginning of the iteration are not necessary to explain the value
withdrawal. It is possible to explain the value withdrawal by a subset
of these rules such that every iteration using this subset of rules
removes the considered value. This subset of rules is an explanation
of the value withdrawal.
This notion of explanation is declarative (does not depend on the
computation).
We are going to define a more precise notion
of explanation: this subset will be structured as a tree.

For the sake of clarity, it will be achieved first in a basic but
significant full arc consistency like framework. Next, we extend it to
weaker arc consistencies (partial arc consistency), and finally to
a framework including hyper-arc consistency as a special case.
The full and the partial hyper-arc consistency of GNU-Prolog are
instances of this framework.

\medskip

First, we consider special reduction rules called rules of ``abstract arc
consistency''. Such a rule is binary and its type has the form
$(\{x,y\}, y)$, that is it reduces the domain of $y$ using the domain
of $x$.

An {\em abstract arc consistency reduction rule} $r$ is defined by two
variables $in_r$ and $out_r$ and a function $arc_r : D_{out_r}
\rightarrow {\cal P}(D_{in_r})$.

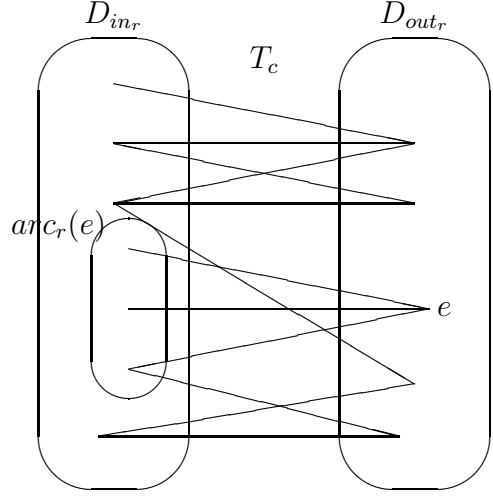
\begin{figure}[t]
\begin{center}
\setlength{\unitlength}{4144sp}%
\begingroup\makeatletter\ifx\SetFigFont\undefined%
\gdef\SetFigFont#1#2#3#4#5{%
  \reset@font\fontsize{#1}{#2pt}%
  \fontfamily{#3}\fontseries{#4}\fontshape{#5}%
  \selectfont}%
\fi\endgroup%
\begin{picture}(2857,2932)(752,-2768)
\thinlines
\put(1351,-1411){\oval(900,2700)}
\put(3151,-1411){\oval(900,2700)}
\put(1441,-1681){\oval(450,1080)}
\put(1351,-331){\line( 5,-1){1800}}
\put(3151,-691){\line(-1, 0){1800}}
\put(3151,-691){\line(-5,-1){1800}}
\put(3151,-1051){\line(-1, 0){1800}}
\put(3151,-1051){\line(-5, 1){1800}}
\put(1441,-1321){\line( 5,-1){1800}}
\put(3241,-1681){\line(-1, 0){1800}}
\put(3241,-1681){\line(-5,-1){1800}}
\put(1351,-1051){\line( 5,-3){1800}}
\put(3151,-2131){\line(-6,-1){1890}}
\put(1441,-2041){\line( 4,-1){1620}}
\put(3061,-2446){\line(-1, 0){1800}}
\put(1351, 29){\makebox(0,0)[b]{\smash{\SetFigFont{12}{14.4}{\rmdefault}{\mddefault}{\updefault}
$D_{in_r}$
}}}
\put(3151, 29){\makebox(0,0)[b]{\smash{\SetFigFont{12}{14.4}{\rmdefault}{\mddefault}{\updefault}
$D_{out_r}$
}}}
\put(2251,-241){\makebox(0,0)[b]{\smash{\SetFigFont{12}{14.4}{\rmdefault}{\mddefault}{\updefault}
$T_c$
}}}
\put(3286,-1726){\makebox(0,0)[lb]{\smash{\SetFigFont{12}{14.4}{\rmdefault}{\mddefault}{\updefault}
$e$
}}}
\put(1306,-1231){\makebox(0,0)[rb]{\smash{\SetFigFont{12}{14.4}{\rmdefault}{\mddefault}{\updefault}
$arc_r(e)$
}}}
\end{picture}
\end{center}
\caption{The particular case of arc consistency}
\label{Figure:The particular case of arc consistency}
\end{figure}

Intuitively, $out_r$ is the variable whose domain is modified according
to the domain of the other variable $in_r$, and, for $e \in
D_{out_r}$, $arc_r(e)$ is a superset of the values connected to $e$
by the constraint associated with $r$ (see for example
figure~\ref{Figure:The particular case of arc consistency}).

Formally, the type of $r$ is $(\{ in_r, out_r\}, out_r)$ and $r(d)=\{
e \in d_{out_r} \mid arc_r(e) \cap d_{in_r} \not= \emptyset\}$ for
each $d \in {\cal D}(\{in_r, out_r\})$.

So we have the obvious lemma:

\begin{lemma}
For each abstract arc consistency reduction rule $r$ and $e \in
D_{out_r}$, and for each $d \in {\cal D}(\{in_r, out_r\})$
\[
\left( \bigwedge_{f \in arc_r(e)} f \not\in d_{in_r} \right)
\Rightarrow e \not\in r(d)
\]
\end{lemma}

In particular, if $arc_r(e) = \emptyset$ then we have $e \not\in r(d)$.

\begin{example}{Arc consistency}
In the framework of arc consistency, each constraint $c$ is binary,
that is $\var(c)=\{x,y\}$, and it provides two rules: $r_1$ of type
$(\{x,y\},x)$, $r_1(d) = \{ e \in d_x \mid \exists f \in d_y, (x
\mapsto e, y \mapsto f) \in T_c\}$, that is, for each $e \in D_x$,
$arc_{r_1}(e) = \{f \in D_y \mid (x \mapsto e, y \mapsto f) \in
T_c\}$, and the other rule $r_2$ of type $(\{x,y\},y)$ defined
similarly.

Note that it is possible to define weaker notions of arc
consistency, such that $arc_{r_1}(e) \supseteq \{f \in D_y \mid (x
\mapsto e, y \mapsto f) \in T_c\}$. But, it will be dealt later in a more
general framework.
\end{example}

\begin{example}{GNU-Prolog}
Let us consider the constraint ``{\tt x~\#<~y}'' in GNU-Prolog. 
This constraint is implemented by two reduction rules, it is the {\em
glass-box} paradigm \cite{CodDia-jlp-96,DevSarVan-draft-91}:
\begin{enumerate}
\item $r_1$ of type $(\{x,y\},x)$ (i.e. $in_{r_1}=y$, $out_{r_1}=x$),
with, for each $e \in D_x$, $arc_{r_1}(e)=\{f \in D_y \mid e < f\}$;
\item $r_2$ of type $(\{x,y\},y)$ (i.e. $in_{r_2}=x$, $out_{r_2}=y$),
with, for each $e \in D_y$, $arc_{r_2}(e)=\{f \in D_x \mid f < e\}$.
\end{enumerate}
\end{example}

Let us suppose that each $r \in R$ is such an abstract arc consistency
reduction rule.

Let us consider an iteration $d^0, d^1, \ldots$ of $R$ from $D$ with
respect to the run $r_1, r_2, \ldots$.  Let us assume that the value
$e$ has disappeared from the domain of the variable $out_{r_i}$ at the
$i$-th step, that is $e \in d^{i-1}_{out_{r_i}}$ but $e \not\in
d^i_{out_{r_i}}$. Note that $d^i_{out_{r_i}} =
r_i(d^{i-1}|_{\{in_{r_i},out_{r_i}\}}) = \{ e \in d^{i-1}_{out_{r_i}}
\mid arc_{r_i}(e) \cap d^{i-1}_{in_{r_i}} \not= \emptyset \}$. So
$arc_{r_i}(e) \cap d^{i-1}_{in_{r_i}} = \emptyset$, i.e. for each $f
\in arc_{r_i}(e)$, $f \not\in d^{i-1}_{in_{r_i}}$.

According to the previous lemma $e \not\in d^{i}_{out_{r_i}}$ because
$\bigwedge_{f \in arc_{r_i}(e)} f \not\in d^{i-1}_{in_{r_i}}$.
But if $f \not\in d^{i-1}_{in_{r_i}}$ it is because there exists $j_f
< i$ such that $f$ has disappeared at the $j_f$-th step that is $f \in
d^{j_{f}-1}_{in_{r_i}}$ but $f \not\in d^{j_{f}}_{in_{r_i}}$
(note that $in_{r_i} = out_{r_{j_f}}$).

Let us define $p(e,i)=\{(f,j) \mid f \in arc_{r_i}(e), f \not\in
d^{j}_{out_{r_j}}, f \in d^{j-1}_{out_{r_j}}\}$.

So $e \not\in d^i_{out_{r_i}}$ because $\bigwedge_{(f,j) \in p(e,i)} f \not\in d^{j}_{out_{r_j}}$.

We are going to define the notion of explanation by {\em abstracting} $d$:

\begin{definition} (compare with the previous lemma)
For each reduction rule $r$, for each $e \in D_{out_r}$, we define
the {\em deduction rule} named $(e, r)$:
\[
(e,r) : (e, out_r) \leftarrow \{ (f, in_r) \mid f \in arc_r(e) \}
\]
$(e, out_r)$ is the head of the rule and $\{ (f, in_r) \mid f \in
arc_r(e) \}$ is its body.
\end{definition}

In particular, when $arc_r(e)=\emptyset$ the body is empty and the
deduction rule is reduced to ``the fact'' $(e,r) : (e,out(r))
\leftarrow \emptyset$.

Intuitively, a deduction rule $(e,r) : (e, out(r)) \leftarrow \{ (f,
in_{r}) \mid f \in arc_r(e) \}$ should be understood as follow: if all
the $f \in arc_r(e)$ are removed from the domain of $in_r$ then $e$
is removed from the domain of $out(r)$.

The set of deduction rules $(e,r)$ where $r \in R$ and $e \in
D_{out_r}$ is exactly an {\em inductive definition} according to
\cite{Aczel-handbook-77}, and a proof tree rooted by $(e,y)$ where
$y = out_r \in V$ and $e \in D_y$ will be called an explanation for $(e,y)$.
Note that a leaf of an explanation corresponds to a fact $(e,r)$ that
is the case where $arc_r(e) = \emptyset$.

Intuitively, the proof tree provides an explanation of the reason why
$e$ may be removed from the domain of $y$.

It is important to note that an explanation is merely a tree made of
deduction rules, i.e. the $d^i$ of an iteration are not part of the
explanation.

\begin{example}{GNU-Prolog}\label{Example:explanation}
Let us consider the 3 constraints {\tt x~\#<~y}, {\tt y~\#<~z}, {\tt
z~\#<~x} with $D_x = D_y = D_z = \{ 0, 1, 2 \}$. The reduction rules
are: 
\begin{itemize}
\item $r_1$ of type $(\{x,y\}, x)$, defined by $r_1(d) = \{ e \in d_x
\mid arc_{r_1}(e) \cap d_y \neq \emptyset\}$ where $arc_{r_1}(e)=\{f
\in D_y \mid e < f\}$;
\item $r_2$ of type $(\{x,y\}, y)$, defined by $r_2(d) = \{ e \in d_y
\mid arc_{r_2}(e) \cap d_x \neq \emptyset\}$ where $arc_{r_2}(e)=\{f
\in D_x \mid f < e\}$;
\item $r_3$ of type $(\{y,z\}, y)$, defined by $r_3(d) = \{ e \in d_y
\mid arc_{r_3}(e) \cap d_z \neq \emptyset\}$ where $arc_{r_3}(e)=\{f
\in D_z \mid e < f\}$;
\item $r_4$, $r_5$, $r_6$ are defined in the same way.
\end{itemize}

Figure~\ref{Figure:Value Withdrawal Explanations} shows three
different explanations for $(0,x)$.
For example, the first explanation says: $0$ is removed from the
domain of $x$ by the reduction rule $r_1$ because $1$ is removed from
the domain of $y$ and $2$ is removed from the domain of $y$.
$1$ is removed from the domain of $y$ by the reduction rule $r_3$
because $2$ is removed from the domain of $z$, and so on...

The first and third explanations correspond to some iterations
(see example~\ref{Example:explanationsuite}). But the second one does
not correspond to an iteration. This introduces some questions
(which are going to be answered by theorem~\ref{Theorem:First}).

\begin{figure}[t]
\begin{center}
\ptrule{$(0,r_1)$}%
   {$(0,x)$}%
   {\ptrules{$(1,r_3)$}{$(2,r_5)$}
      {$(1,y)$}
      {$(2,z)$}
      {}%
    \ptrule{$(2,r_3)$}%
      {$(2,y)$}%
      {}%
   }
\hspace*{2em}
\ptrule{$(0,r_1)$}%
   {$(0,x)$}%
   {\ptrules{$(1,r_2)$}{$(0,r_6)$}
      {$(1,y)$}
      {$(0,x)$}
      {}%
    \ptrule{$(2,r_3)$}%
      {$(2,y)$}%
      {}%
   }

\ptrule{$(0,r_6)$}{$(0,x)$}{}
\end{center}
\caption{Value Withdrawal Explanations}
\label{Figure:Value Withdrawal Explanations}
\end{figure}
\end{example}

Explanations are very declarative but they can be extracted from
iterations.

We are going to define an explanation associated with the event
``withdrawal of a value from a domain in an iteration''. It is introduced by
the following theorem.

\begin{theorem}\label{Theorem:First}
(There exists an explanation for $(e,y)$) if and only if (there exists
a chaotic iteration with limit $d^{\omega}$ such that $e \not\in
d^{\omega}_y$) if and only if ($e \not\in \Theta_y$, where $\Theta$ is
the downward closure).
\end{theorem}

\begin{proof}
The last equivalence is proved by lemma~\ref{lemma:chaotic iteration}.
About the first one:

$\Leftarrow$: Let $d^0, d^1, ...$ be the chaotic
iteration (with respect to the run $r_1, r_2, \ldots$).
There exists $i$ such that $e \in d^{i-1}_{y}$ but $e
\not\in d^i_y$.

We define a tree $\expl(e, y, i)$ which is an explanation for $(e,y)$.
$\expl(e, y, i)$ is inductively defined as follows:
\begin{itemize}
\item $y = out_{r_i}$;
\item the root of the tree $\expl(e, y, i)$ is labeled by $(e, y)$;
\item (we have previously observed that $e \not\in d^i_{out_{r_i}} = d^i_y$
because $\bigwedge_{(f,j) \in p(e,i)} f \not\in d^{j}_{out_{r_j}}$
and, for each $(f,j) \in p(e,i)$, $out_{r_j} = in_{r_i}$)
the deduction rule used to connect the root to its children,
which are labeled by the $(f,out_{r_j})$, is $(e, r_i) :
(e, out_{r_i}) \leftarrow \{ (f,in_{r_i}) \mid f \in arc_{r_i}(e) \}$;
\item the immediate subtrees of $\expl(e, y, i)$ are the
$\expl(f, in_{r_i}, j)$ for $(f,j) \in p(e,i)$.
\end{itemize}

$\Rightarrow$:
let us consider a numbering $1, \ldots, n$ of the nodes of the
explanation such that the traversal according to the numbering from
$n$ to $1$ corresponds to a breadth first search algorithm.  For each
$i \in \{1, \ldots, n\}$, let $(e_i, r_i)$ be the name of the rule
which links the node $i$ to its children, and let $d^0, \ldots, d^n$
be the prefix of every iteration w.r.t. a run which starts by $r_1,
\ldots, r_n$. By induction we show that $e_i \not\in d^i_{out(r_i)}$,
so $e \not\in d^{\omega}_y$ for every iteration whose run starts by
$r_1, \ldots, r_n$.
\end{proof}

It is important to note that the previous proof is constructive.  The
definition of $\expl(e, y, i)$ gives an incremental algorithm to
compute explanations.

\begin{example}{GNU-Prolog}\label{Example:explanationsuite}
(Continuation of example~\ref{Example:explanation})

Let us consider the iteration $r_5$, $r_3$, $r_1$.
The first explanation of figure~\ref{Figure:Value Withdrawal Explanations}
says:

At the beginning, $d^0_x=\{0,1,2\}$.
$arc_{r_5}(2) = \emptyset$ so $2 \not\in d^1_z$.

Then, $d^1_z=\{0,1\}$.
$arc_{r_3}(1) = \{2\}$ so $2 \not\in d^1_z \Rightarrow 1 \not\in d^2_y$.
$arc_{r_3}(2) = \emptyset$ so $2 \not \in d^2_y$.

Then, $d^2_y=\{0\}$.
$arc_{r_1}(0) = \{1,2\}$ so $(1 \not\in d^2_y \wedge 2 \not\in d^2_y)
\Rightarrow 0 \not\in d^3_x$.
\end{example}

We extend our formalization in order to include weaker arc consistency
rules.
In GNU-Prolog, a full arc consistency rule uses the whole domain of
the input variable, whereas, a partial arc consistency rule only uses
its lower and upper bounds. In that case we need two functions $arc$,
one for each bound.

An {\em abstract arc consistency reduction rule} $r$ is now defined by
two variables $in_r$ and $out_r$ and a set $Arc_r$ of functions
$D_{out_r} \rightarrow {\cal P}(D_{in_r})$.

The type of $r$ is $(\{in_r, out_r\}, out_r)$ and
$r(d)=\{ e \in d_{out_r} \mid \bigwedge_{arc \in Arc_r} (arc(e) \cap
d_{in_r} \not= \emptyset)\}$ for each
$d \in {\cal D}(\{in_r, out_r\})$.

Note that for arc consistency, $Arc_r$ contains only one function
(it is the previous framework).

Obviously, for each $arc \in Arc_r$ we have:
\[
\left( \bigwedge_{f \in arc(e)} f \not\in d_{in_r} \right)
\Rightarrow e \not\in r(d)
\]

\begin{example}{GNU-Prolog}
Let us consider the constraint ``{\tt x~\#=~y+c}'' in GNU-Prolog where
$x$, $y$ are variables and $c$ a constant. This constraint is
implemented by two reduction rules: $r_1$ of type $(\{x,y\},x)$
(i.e. $in_{r_1}=y$, $out_{r_1}=x$) and $r_2$ of type $(\{x,y\},y)$. In
GNU-Prolog, $r_1$ is defined by the partial arc consistency rule
{\tt x~in~min(y)+c..max(y)+c}.

$r_1(d)=\{e \in d_x \mid arc_{r_1}^1(e) \cap d_y \neq \emptyset \wedge
arc_{r_1}^2(e) \cap d_y \neq \emptyset\}$ where
$arc_{r_1}^1(e)=\{f \in D_y \mid f+c \leq e\}$ and
$arc_{r_1}^2(e)=\{f \in D_y \mid e \leq f+c\}$.

$r_2$ of type $(\{x,y\},y)$ is defined in the same way by the
rule {\tt y~in~min(x)-c..max(x)-c}.
\end{example}

Let us suppose that each $r \in R$ is such  an abstract arc consistency
reduction rule.

\begin{definition}
For each reduction rule $r$, for each $e \in D_{out_r}$, for each
$arc \in Arc_r$, we define the {\em deduction rule} named $(e, r, arc)$:
\[
(e, r, arc) : (e, out_r) \leftarrow \{ (f, in_r) \mid f \in arc(e) \}
\]
\end{definition}

Again the set of deduction rules $(e, r, arc)$ is an inductive
definition and this defines a generalization of our previous notion of
explanation.

\medskip

Now we generalize the reduction rules to {\em hyper-arc consistency}.

An {\em abstract hyper-arc consistency reduction rule} $r$ is defined by
{\em a set} of variables $in_r$, a variable $out_r$ and a set $Arc_r$
of functions $D_{out_r} \rightarrow {\cal P}(\prod_{x \in in_r} D_x)$.

The type of $r$ is $(in_r \cup \{out_r\}, out_r)$ and
$r(d)=\{ e \in d_{out_r} \mid \bigwedge_{arc \in Arc_r}
(arc(e) \cap \prod_{x \in in_r} d_x \not= \emptyset)\}$
for each $d \in {\cal D}(in_r \cup \{out_r\})$.

Note that for hyper-arc consistency, $Arc_r$ contains only one function.

Obviously, for each $arc \in Arc_r$ we have:
\[
\left( \bigwedge_{f \in arc(e)} (f \not\in \prod_{x \in in_r} d_x) \right)
\Rightarrow e \not\in r(d)
\]

that is

\[
\left( \bigwedge_{f \in arc(e)} \left( \bigvee_{x \in in_r}
f_x \not\in d_x \right) \right) \Rightarrow e \not \in r(d)
\hspace*{1cm} \mbox{\large (1)} \hspace*{-1.5cm}
\]

But
\[
\left( \bigwedge_{f \in arc(e)} \left( \bigvee_{x \in in_r}
f_x \not\in d_x \right) \right)
\Leftrightarrow
\left( \bigvee_{t : arc(e) \rightarrow in_r} \left( \bigwedge_{f \in arc(e)}
f_{t(f)} \not\in d_{t(f)} \right) \right)
\hspace*{1cm} \mbox{\large (2)} \hspace*{-1.5cm}
\]
Intuitively, the $t : arc(e) \rightarrow in_r$ are choice functions
and each $t(f)$ is one $x$ such that $f_x \not\in d_x$.

So we have, for each $t : arc(e) \rightarrow in_r$,
\[
\left( \bigwedge_{f \in arc(e)} f_{t(f)} \not\in d_{t(f)} \right)
\Rightarrow e \not \in r(d)
\hspace*{1cm} \mbox{\large (3)} \hspace*{-1.5cm}
\]

\begin{example}{Hyper-arc Consistency in GNU-Prolog}
Let us consider the constraint ``{\tt x~\#=\#~y+z}'' in GNU-Prolog.
Let $D_x=D_y=D_z=\{1,2,3\}$.
The constraint is implemented by three reduction rules:
\begin{itemize}
\item $r_1$ of type $(\{x,y,z\},x)$ defined by
$r_1(d) = \{e \in d_x \mid \exists f \in \prod_{v \in \{y,z\}} d_v,
e = f_y + f_z\}$;
\item $r_2$ of type $(\{x,y,z\},y)$ and $r_3$ of type $(\{x,y,z\},z)$
defined in the same way.
\end{itemize}
Here, $in_{r_1} = \{y, z \}$, $out_{r_1} = x$ and $Arc_{r_1} = \{
arc \}$, where
$arc(e)=\{f \in \prod_{v \in \{y,z\}} D_{v} \mid
e = f_y + f_z \}$.

For example, $arc(3) = \{ (y \mapsto 1, z \mapsto 2),
(y \mapsto 2, z \mapsto 1) \}$.
We have as an instance of $(1)$:
$(1 \not\in d_y \vee 2 \not\in d_z) \wedge (2 \not\in d_y \vee 1
\not\in d_z) \Rightarrow 3 \not\in r_1(d)$.
But $(1 \not\in d_y \vee 2 \not\in d_z) \wedge (2 \not\in d_y \vee 1
\not\in d_z)$ is equivalent to
$(1 \not\in d_y \wedge 2 \not\in d_y) \vee
(1 \not\in d_y \wedge 1 \not\in d_z) \vee
(2 \not\in d_z \wedge 2 \not\in d_y) \vee
(2 \not\in d_z \wedge 1 \not\in d_z)$
This equivalence is the corresponding instance of $(2)$.
So we have the following instances of $(3)$:
\begin{itemize}
\item $1 \not \in d_y \wedge 2 \not \in d_y \Rightarrow 3 \not \in
r_1(d)$;
\item $1 \not \in d_y \wedge 1 \not \in d_z \Rightarrow 3 \not \in
r_1(d)$;
\item $2 \not \in d_z \wedge 2 \not \in d_y \Rightarrow 3 \not \in
r_1(d)$;
\item $2 \not \in d_z \wedge 1 \not \in d_z \Rightarrow 3 \not \in
r_1(d)$.
\end{itemize}
\end{example}

Again a more general notion of explanation is defined by abstracting
$d$.

Let us suppose that each $r \in R$ is such an abstract hyper-arc
consistency reduction rule.

\begin{definition}
For each reduction rule $r$, for each $e \in D_{out_r}$, for each
$arc \in Arc_r$, for each $t : arc(e) \rightarrow in_r$,
we define the {\em deduction rule} named $(e, r, arc, t)$:
\[
(e, r, arc, t) : (e, out_r) \leftarrow \{ (f_{t(f)}, t(f))
\mid f \in arc(e) \}
\]
\end{definition}

The inductive definition for hyper-arc consistency is larger than for arc
consistency because of the number of variables of the rules, but in
practice (GNU-Prolog), the rules contain two or three variables.

\section{Conclusion}
\label{Sect:Conclusion}

This paper has given a model for the operational semantics of CSP
solvers by domain reduction.

This model is applied to the definition of a notion of explanation.
An explanation is a set of rules structured as a tree. An interesting
aspect of our definition is that a subtree of an explanation is also
an explanation (inductive definition).

This model can be applied to usual constraint solvers using
propagation, for example it takes into account the full and the
partial hyper-arc consistency of GNU-Prolog.

As it is written in the introduction, constraint solving combines
domain reduction and labeling. A perspective is to really take into
account labeling in our model.

We plan to use explanations in order to diagnose errors in a CSP
(according to an expected semantics), in the style of
\cite{Shapiro-phd-82,Ferrand-aadebug-93}.

Another perspective is to take advantage of the glass-box model
\cite{CodDia-jlp-96} and more generally the S-box model
\cite{GouBen-ase-99} and to distinguish different levels of rules
({\tt x in r}, built-in constraints, S-box, ...)

\section*{Acknowledgements}
Discussions with Patrice Boizumault and Narendra Jussien are
gracefully acknowledged.
We would also like to thank our anonymous referees for the helpful comments.

\end{document}